\documentclass[12pt]{article}
\usepackage{epsfig}
\usepackage{graphicx}

\newcommand{\dd}{\mbox{d}}
\newcommand{\eps}{\varepsilon}
\newcommand{\Li}{\mbox{Li}_2}
\newcommand{\vecc}[1]{\mbox{\boldmath $#1$}}
\newcommand{\matrm}[1]{{\mbox{\scriptsize #1}}}

\def\fun#1#2{\lower3.6pt\vbox{\baselineskip0pt\lineskip.9pt
\ialign{$\mathsurround=0pt#1\hfil##\hfil$\crcr#2\crcr\sim\crcr}}}

\title{Tagged photons in DIS with next-to-leading accuracy}

\author{H.~Anlauf%
\thanks{Supported by Bundesministerium f\"ur Bildung, Wissenschaft,
        Forschung und Technologie (BMBF), Germany.}$\;^{,1}$,
A.B.~Arbuzov$^{2}$, E.A.~Kuraev$^{2}$, N.P.~Merenkov$^{3}$}

\date{}

\begin{document}

\maketitle

\begin{center}
{$^1$ \it Fachbereich Physik, Siegen University,
57068 Siegen, Germany}\\[.2cm]
{$^2$ \it Bogoliubov Laboratory of Theoretical Physics, JINR, \\
Dubna, 141980, Russia} \\[.2cm]
{$^3$ \it Kharkov Institute of Physics and Technology, \\
Kharkov, 310108, Ukraine} \\[.2cm]
\end{center}

\begin{abstract}
The leading and next-to-leading radiative corrections to deep inelastic
events with tagged photons are calculated analytically.  Comparisons
with previous results and numerical estimations are presented for the
experimental conditions at HERA.
\end{abstract}

\section{Introduction}

It is well known that the radiative corrections to deep inelastic
electron proton scattering due to hard real photon emission are very
important in certain regions of the HERA kinematic domain.  In fact, the
initial-state collinear radiation leads to a reduction of the projectile
electron energy and therefore to a shift of the effective Bjorken
variables in the hard scattering process as compared to those determined
from the actual measurement of the scattered electron alone.  Therefore,
radiative events
\begin{equation}\label{1}
e(p_1) + p(P) \rightarrow e(p_2) + \gamma(k) + X + (\gamma)
\end{equation}
are to be carefully taken into account \cite{mo,heraws:1991,mi}.

On the other hand, measuring the energy of the photons emitted very
close to the incident electron beam direction \cite{jad,kra,bar,H1:rad}
permits to overlap the kinematical region of photoproduction
$(Q^2\approx 0)$ and the DIS region with small transferred momenta
(about a few GeV$^2$) within the high energy HERA experiments.
Furthermore, these radiative events may be used to independently
determine the proton structure functions $F_2$ and $F_1$ (and therefore
$F_L$) in a single run without lowering the beam energies
\cite{kra,FGMZ96}.
Preliminary results of an $F_2$ analysis using such radiative events were
recently presented by the H1 collaboration \cite{H1:ISR-prelim}.

Our aim is to calculate the radiative corrections to neutral current
deep inelastic events with simultaneous (exclusive) detection of a hard photon
emitted very close to the direction of the incoming electron beam
$(\theta_\gamma = \widehat{\vecc{p}_1\vecc{k}} \leq \theta_0 \approx
5\cdot 10^{-4}$~rad).  In the case of the HERA collider, the
experimental detection of photons emitted in this very forward direction
is actually possible due to the presence of photon detectors (PD) that
are part of the luminosity monitoring system of ZEUS and H1.

Let us briefly review the kinematics for the process under
consideration.  As the opening angle of the forward photon detector is
very small, and since we will only consider cross sections where the
tagged photon is integrated over the solid angle covered by this photon
detector, we can parameterize these radiative events using the
standard Bjorken variables $x$ and $y$, that are determined from the 
measurement of the scattered electron,
\begin{eqnarray}
        x &=& \frac{Q^2}{2 P\cdot (p_1 - p_2)}\,, \quad
        y = \frac{2 P\cdot (p_1 - p_2)}{V}\,, \nonumber\\
        Q^2 &=& 2 p_1\cdot  p_2 = xyV, \quad
        \mbox{with} \quad V = 2 P\cdot  p_1\,,
\end{eqnarray}
and the energy fraction $z$ of the electron after initial state
radiation of a collinear photon,
\begin{equation}
        z = \frac{2P\cdot  (p_1 - k)}{V} = \frac{\eps - k^0}{\eps}\,,
\end{equation}
where $\eps$ is the initial electron energy, and $k^0$ is the energy seen
in the forward photon detector.

An alternative set of kinematic variables that is especially adapted to
the case of collinear radiation, is given by the {\em shifted} Bjorken
variables \cite{kra}
\begin{equation}
\label{eq:kin-vars}
        \hat{Q}^2 = -(p_1-p_2-k)^2\, ,
	\qquad
        \hat{x} = \frac{\hat{Q}^2}{2P\cdot  (p_1-p_2-k)}\,,
	\qquad
        \hat{y} = \frac{P\cdot  (p_1-p_2-k)}{P\cdot  (p_1-k)}\, .
\end{equation}
The relations between the shifted and the standard Bjorken variables
read \cite{kra}
\begin{equation}
\label{4}
	\hat{Q}^2   = zQ^2\,, \qquad
	\hat{x}     = \frac{xyz}{z + y - 1}\,, \qquad
	\hat{y}     = \frac{z + y - 1}{z}\,.
\end{equation}

The cross-section under consideration in the Born
approximation,
integrated over the solid angle of the photon detector
($0 \leq \theta_\gamma \leq \theta_0, \; \theta_0 \ll 1$) then takes the
following form:
\begin{equation} \label{Born}
\frac{z}{y} \frac{\dd^3\sigma_\mathrm{Born}}{\dd x \, \dd y \, \dd z} =
\frac{1}{\hat{y}}
\frac{\dd^3\sigma_\mathrm{Born}}{\dd\hat{x}\,\dd\hat{y}\,\dd z} =
\frac{\alpha}{2\pi}P(z,L_0) \tilde\Sigma\,,
\end{equation}
%
where
$$\tilde\Sigma=\Sigma(\hat{x},\hat{y},\hat{Q}^2) = \frac{2\pi\alpha^2(-\hat{Q}^2)}
{\hat{Q}^2\hat{x}\hat{y}^2} F_2(\hat{x},\hat{Q}^2)
\biggl[ 2(1-\hat{y}) - 2\hat{x}^2\hat{y}^2\frac{M^2}{\hat{Q}^2}
+ \biggl(1+4\hat{x}^2\frac{M^2}{\hat{Q}^2}\biggr) \frac{\hat{y}^2}{1+R}
\biggr]\,,$$
$$
P(z,L_0) = \frac{1+z^2}{1-z}L_0 - \frac{2z}{1-z}\,, \quad
R = R(\hat{x},\hat{Q}^2) =
\biggl(1+4\hat{x}^2\frac{M^2}{\hat{Q}^2}\biggr)
\frac{F_2(\hat{x},\hat{Q}^2)}{2\hat{x}F_1(\hat{x},\hat{Q}^2)}\,, - 1
\,,
$$
$$\alpha(-\hat{Q}^2) = \frac{\alpha}{1-\Pi(-\hat{Q}^2)}\,, \quad
L_0  = \ln\left(\frac{\varepsilon^2\theta_0^2}{m^2}\right)\,, \quad
\hat{Q}^2 = 2zp_1\cdot p_2 = 2z\varepsilon^2Y (1-c)\,,$$
$$Y   = \frac{\varepsilon_2}{\varepsilon}=1-y
+ xy \frac{E_p(1+\beta_p)}{2\varepsilon}\,, \quad
c    = \cos(\widehat{\vecc{p}_1\vecc{p}}_2)\,,$$
$$\hat{x} = \frac{\hat{Q}^2}{2P\cdot (zp_1-p_2)}
         = \frac{z\varepsilon Y(1-c)}{zE_p(1+\beta_p)-YE_p(1+\beta_pc)}\,, \quad
\beta_p = \sqrt{1-M^2/E_p^2}\,,$$
\begin{equation}\label{2}
\hat{y} = \frac{2P\cdot (zp_1-p_2)}{zV}
        = \frac{z(1+\beta_p)-Y(1+\beta_pc)}{z(1+\beta_p)}\,.
\end{equation}

The quantities $F_2$ and $F_1$ are the proton structure functions;
$M$ and $m$ are the proton and electron masses, respectively.
In the cross-section~(\ref{Born}) we take
into account terms proportional to $M^2/\hat{Q}^2$\,,
which may be important at low $Q^2$.
Note that the neglect of $Z$-boson exchange and $\gamma$--$Z$
interference is a good approximation, because we are interested
mostly in events with small momentum transfer $\hat{Q}^2$.%
\footnote{The corresponding Born cross section including contributions
from the $Z$ can be found in ref.~\cite{bar}.}
The energies of the initial and final electron, of the tagged photon and
of the initial proton $(\varepsilon$, $\varepsilon_2$, $k^0$ and $E_p)$
are defined in the laboratory reference frame (i.e., the rest frame of
HERA detectors).  The cross section~(\ref{Born}) agrees
with~\cite{kra,bar}.  Also note that we explicitly included the
correction from the vacuum polarization operator $\Pi(-\hat{Q}^2)$ in
the virtual photon propagator.  The aim of our work is to calculate the
higher order QED radiative corrections for this process in the leading
and next-to-leading logarithmic approximation.

In this paper we restrict ourselves to the model independent QED
radiative corrections related to the lepton line, which form a complete, gauge
invariant subset for the neutral current scattering process.  The
remaining source of QED radiative corrections at the same order, such as
virtual corrections with double photon exchange and bremsstrahlung off
the partons are more involved and model dependent, they will be considered
elsewhere.  Our
approach to the calculation of the QED corrections is based on the
utilization of all essential Feynman diagrams that describe the observed
cross-section in the framework of the used approximation.  The same
approach was used recently for the calculation of the QED corrections
for the small angle Bhabha scattering cross-section at
LEP1~\cite{bhabha}.  This work extends our previous calculation at the
leading logarithmic level \cite{AAKM:ll} and presents the details of the
brief outline published in \cite{AAKM:JETP}.

The paper is organized as follows.  Section~\ref{sez2} is devoted to the
corrections related with emission of virtual and soft real photons in
the hard collinear photon emission DIS process.  In sect.~\ref{sez3} we
consider the radiative corrections due to emission of two hard photons
in the collinear kinematics (where we distinguish between the cases when
both photons are emitted close to the initial electron direction and the
case when one of the photons is emitted along the initial and the other
one along the scattered electron direction) and the semi-collinear
kinematics, where the additional hard photon is emitted at a large
angle.  Section~\ref{sez4} collects the results obtained and discusses two
experimental cases: an exclusive set-up, that assumes that a bare
electron can be measured, and a calorimetric one.  We show that, in the
latter case for a coarse detector that performs a calorimetric
measurement, we can reproduce the result of our previous paper
\cite{AAKM:ll}, where the leading logarithmic approximation was used and
where the emission along the final electron was not taken into account.
In conclusion we give some numerical estimates.  The appendices are
devoted to details of the calculation.

\section{Virtual and soft corrections}
\label{sez2}

In order to calculate the contributions from the virtual and soft photon
emission corrections, we start from the expression for the Compton
scattering tensor with a heavy photon~\cite{compt},
\begin{equation}\label{5}
K_{\mu\nu}=(8\pi\alpha)^{-1}\sum_{\mathrm{spins}}
M_{\mu}^{e\gamma^*\to e'\gamma}
(M_{\nu}^{e\gamma^*\to e'\gamma})^*\,,
\end{equation}
where $M_{\mu}$ is the matrix element of the process of Compton scattering
\begin{equation}\label{6}
\gamma^*(-q)+e(p_1) \to \gamma(k)+e(p_2)\,,
\end{equation}
and the index $\mu$ describes the polarization state of the virtual
photon.  This tensor is conveniently decomposed as follows:
\begin{eqnarray}\label{7}
K_{\mu\nu} &=& \frac{1}{2} ( P_{\mu\nu} + P_{\nu\mu}^{*})\,, \\ \nonumber
P_{\mu\nu}
&=& \tilde{g}_{\mu\nu}(B_g + \frac{\alpha}{2\pi}T_g)
+ \tilde{p}_{1\mu}\tilde{p}_{1\nu}(B_{11} + \frac{\alpha}{2\pi}T_{11}) +
\\ \nonumber
&&+ \tilde{p}_{2\mu}\tilde{p}_{2\nu}(B_{22} + \frac{\alpha}{2\pi}T_{22})
+ \frac{\alpha}{2\pi}( \tilde{p}_{1\mu}\tilde{p}_{2\nu}T_{12}
+ \tilde{p}_{2\mu}\tilde{p}_{1\nu}T_{21} ), \\ \nonumber
\tilde{g}_{\mu\nu} &=& g_{\mu\nu} - \frac{q_{\mu}q_{\nu}}{q^2}\,,
\quad
\tilde{p}_{1\mu} = p_{1\mu} - q_{\mu}\frac{p_1\cdot q}{q^2}\, , \quad
\tilde{p}_{2\mu} = p_{2\mu} - q_{\mu}\frac{p_2\cdot  q}{q^2}\, , \quad
p_1 = q + p_2 + k\,.
\end{eqnarray}
The expressions for the quantities $B_{ij}$ corresponding to the Born
approximation are:%
\footnote{We have already dropped those terms that vanish in the
high-energy limit when one integrates over any finite region of photon
phase space.}
\begin{eqnarray}\label{8}
B_g    &=& \frac{1}{st}\left[(s+u)^2 + (t+u)^2\right]
- 2m^2q^2\left(\frac{1}{s^2} + \frac{1}{t^2}\right), \qquad
B_{11} = \frac{4q^2}{st} - \frac{8m^2}{s^2}\, ,\nonumber \\
B_{22} &=& \frac{4q^2}{st} - \frac{8m^2}{t^2}\, , \qquad
s = 2p_2\cdot k\,, \quad t = - 2p_1\cdot k, \quad u = (p_2 - p_1)^2\,, \nonumber \\
q^2 &=& s+t+u, \quad p_1^2=p_2^2=m^2,\quad k^2 = 0\,.
\end{eqnarray}
The one-loop QED corrections are contained in the quantities $T_{ij}$,
whose explicit expressions are given in~\cite{compt}.  Here we have to
integrate them over the solid angle of the emitted photon corresponding
to the shape of the photon detector.  We need to keep only the terms
singular in the limit $\theta_\gamma\to 0$, since after integration the
constant terms contribute only proportional to $\theta_0^2 \sim 10^{-6}$
and can be safely neglected.  Another simplification comes from the fact
that we need only the symmetric (and real) part of the tensor $K$.  This
way, by using typical integrals
\begin{eqnarray}\label{9}
\int\frac{\dd \Omega_k}{2\pi}\frac{1}{t}
 = - \frac{L_0}{2\varepsilon^2(1-z)}\,, \qquad
\int\frac{\dd \Omega_k}{2\pi}\frac{m^2}{t^2}
 = \frac{1}{2\varepsilon^2(1-z)^2}\, ,
\end{eqnarray}
and using the expressions given in Appendix~A
we obtain the following expression for the
Compton tensor integrated over the angular part of the photon phase space:
\begin{eqnarray} \label{10}
\int\frac{\dd \Omega_k}{2\pi}K_{\mu\nu} &=& \left(-Q_l^2g_{\mu\nu}
+ 4zp_{1\mu}p_{1\nu} \right)\frac{1}{2\varepsilon^2(1-z)}
\biggl[\biggl(1+\frac{\alpha}{2\pi}\rho\biggr)P(z,L_0)
- \frac{\alpha}{2\pi}T \biggr], \\ \nonumber
\rho &=& 4\ln\frac{\lambda}{m}(L_Q-1)-L^2_Q+3L_Q+3\ln z+\frac{\pi^2}{3}
-\frac{9}{2}\, ,\\ \nonumber
T &=&  \frac{1+z^2}{1-z}(A\ln z + B)
- \frac{4z}{1-z}L_Q\ln z
- \frac{2-(1-z)^2}{2(1-z)}L_0 + {\mathcal O}({\mathrm{const}})\,,   \\ \nonumber
\quad A &=&-L_0^2+2L_0L_Q-2L_0\ln(1-z),
\quad B = \left(\ln^2 z - 2 \Li (1-z)\right)L_0\,,\quad  \\ \nonumber
L_Q & = & \ln\frac{Q^2}{m^2}\,, \qquad
\Li (x) = -\int\limits_{0}^{x}\frac{\dd y}{y}\ln(1-y) \, .
\end{eqnarray}
The quantity $\lambda$, which enters into the expression for $\rho$, is a
fictitious photon mass.

In the construction of the total expression for the tensor $K_{\mu\nu}$
we replaced $q_\mu=q_\nu=0$, $p_{2\mu,\nu}=zp_{1\mu,\nu}$, bearing in
mind the gauge invariance of hadronic tensor \cite{feyn},
\begin{eqnarray}\label{11}
H_{\mu\nu} &=& \frac{4\pi}{M}\left(W_2(x_h,Q^2_h)\tilde P_{\mu}
\tilde P_{\nu}-M^2W_1(x_h,Q^2_h)\tilde g_{\mu \nu}\right), \quad
x_h=\frac{Q^2_h}{2P\cdot q_h}\,, \\ \nonumber
\tilde P_{\nu} &=& P_{\nu}-q_{h\nu}\frac{P\cdot q_h}{q_h^2}\,.
\end{eqnarray}
Here we imply $q_h=q$, $Q^2_h=-q^2$.

Consider now the process with emission of a soft photon
in addition to the emission of the hard one, which hits the PD.
We imply the condidtion that the energy of the soft photon should be
less than some small quantity $\delta\varepsilon$ (in the center--of--mass
system). In straightforward
calculations, starting from Feynman diagrams, some care is to be paid
in the evaluation of integrals over the phase volume of the soft
photon, as some contributions are crucially dependent on the
correlation between our two small parameters $\Delta=
\delta\varepsilon/\varepsilon$ and $\theta_0$.  In our particular case
$\theta_0\ll \Delta \ll 1$, the result coincides with the one obtained
using the approximation of classical currents for soft photons.  The
total effect for the sum of contributions of virtual and soft photon
emission consists in the replacement of the quantity $\rho$ by
$\tilde\rho$ in eq.~(\ref{10}) (see eq.~(45) in \cite{compt}):
\begin{eqnarray}\label{12}
\rho \to \tilde{\rho} = 2(L_Q - 1)\ln\frac{\Delta^2}
{Y} + 3L_Q + 3\ln z - \ln^2Y - \frac{\pi^2}{3} - \frac{9}{2}
+ 2\Li\biggl(\frac{1+c}{2}\biggr)\,.
\end{eqnarray}

The final expression for the virtual and soft photon emission corrected
tagged photon cross-section has the form
\begin{equation}\label{13}
\frac{z}{y} \frac{\dd^3\sigma_{VS}}{\dd x \, \dd y \, \dd z} =
\left(\frac{\alpha}{2\pi}\right)^2 \left[ P(z,L_0)\tilde{\rho}-T \right]
\tilde\Sigma \, .
\end{equation}

\section{Double hard bremsstrahlung}
\label{sez3}

Consider now the emission of an extra photon with the energy more
than $\delta\varepsilon$.
For the calculation of the contributions from real hard bremsstrahlung,
which in our case correspond to double photon emission with at least one
photon seen in the forward detector, we specify three specific
kinematical domains: {\em i)\/} both hard photons strike the forward
photon detector, i.e., both are emitted within a narrow cone around the
electron beam $(\theta \leq \theta_0)$; {\em ii)\/} one hard photon is
tagged by the PD, while the other is collinear to the outgoing electron
$(\theta_2 = \widehat{\vecc{k}_2\vecc{p}}_2 \leq \theta'_0)$; and
finally {\em iii)\/} the second photon is emitted at large angles (i.e.,
outside the defined narrow cones) with respect to both incoming and
outgoing electron momenta.  We denominate the third kinematical domain
as a semi-collinear one.  The contributions of the regions {\em i)\/}
and {\em ii)\/} contain leading terms (quadratic in the large logarithms
$L_0$, $L_Q$), whereas region {\em iii)\/} contains formally non-leading
terms of order $L_0\ln(1/\theta_0^2)$, which, however, give a
contribution numerically larger than the leading ones since $\varepsilon
\theta_0/m \ll 1/\theta_0$.

The calculation beyond the leading logarithmic approximation may be performed
using the results of a paper of one
of us \cite{NPM}.  The contribution from the
kinematical region {\em i)\/} (both hard photons being tagged), has
the form (see eq.~($\Pi$~6) from \cite{NPM}):
\begin{eqnarray} \label{14}
\frac{z}{y} \frac{\dd^3 \sigma^{\gamma\gamma}_i}{\dd x \, \dd y \, \dd z} & = &
\frac{\alpha^2}{8\pi^2}L_0\Biggl[ L_0 \biggl( P^{(2)}_{\Theta}(z)+
2\frac{1+z^2}{1-z}\Bigl(\ln z-\frac{3}{2}-2\ln\Delta\Bigr)\biggr) + 6(1-z)+
\nonumber\\ &&
\phantom{\frac{\alpha^2}{8\pi^2}L_0\Biggl[} 
+ \biggl(\frac{4}{1-z}-1-z\biggr)\ln^2z - 4\frac{(1+z)^2}{1-z}
\ln\frac{1-z}{\Delta}\Biggr]\tilde \Sigma + \nonumber\\
&&\phantom{\frac{\alpha^2}{8\pi^2}L_0\Biggl[} 
+ {\mathcal O}({\mathrm{const}}) \,.
\end{eqnarray}
Here we use the notation $P^{(2)}_{\Theta}(z)$ for the $\Theta$-part of the
second order term of the expansion of the electron non-singlet structure
function
\begin{eqnarray} \label{15}
D(z,L) &=& \delta(1-z)+\frac{\alpha}{2\pi}P^{(1)}(z)L+
\frac{1}{2}\left(\frac{\alpha}{2\pi}\right)^2P^{(2)}(z)L^2 + \dots \,,\nonumber \\
P^{(i)}(z) &=& P^{(i)}_{\Theta}(z)\Theta(1-z-\Delta)+P^{(i)}_{\delta}
\delta(1-z), \quad \Delta\to 0\,, \nonumber \\ 
P^{(1)}_{\Theta}(z) &=& \frac{1+z^2}{1-z}\,, \quad
P^{(1)}_{\delta}=\frac{3}{2}+2\ln\Delta\,, \nonumber\\ 
P^{(2)}_{\Theta}(z) &=& 2\Biggl[\frac{1+z^2}{1-z}\biggl(2\ln(1-z)-\ln z
+\frac{3}{2}\biggr) +\frac{1}{2}(1+z)\ln z-1+z\Biggr]\,.
\end{eqnarray}
The parameter $\Delta$ serves as the infrared regularization parameter.

The contribution of the kinematical region {\em ii)\/} to the observed
cross-section depends on the event selection; in other words, on the
method of measurement of the scattered particles.

In the case of exclusive event selection, when only the scattered
electron is detected, while the photon that is emitted almost
collinearly (i.e., within a small cone with opening angle $2\theta_0'$
around the momentum of the outgoing electron) goes unnoticed or is not
taken into account in the determination of the kinematical variables, we
have (see $\Pi 8$ from \cite{NPM})
\begin{eqnarray}\label{16}
\frac{z}{y}\frac{\dd^3\sigma^{\gamma\gamma}_{ii}}{\dd x \, \dd y \, \dd z} &=&
\frac{\alpha^2}{4\pi^2}P(z,L_0)\int\limits_{\Delta/Y}^{y_2^{\mathrm{max}}}
\frac{\dd y_2}{1+y_2}\biggl[\frac{1+(1+y_2)^2}{y_2}(\widetilde L-1)+y_2
\biggr]\Sigma_s\,, \nonumber\\ 
\Sigma_s &=& \Sigma(x_b,y_b,Q_b^2)\,,
\end{eqnarray}
where
\begin{eqnarray}
\widetilde L &=&\ln(\varepsilon \theta'_0/m)^2+2\ln Y,\quad y_2=\frac{x_2}{Y}\,,
\quad Y=\varepsilon_2/\varepsilon, \quad y_2^{\mathrm{max}}
= \frac{2z-Y(1+c)}{Y(1+c)}\,,\nonumber \\
\label{eq:17}
x_b &=& \frac{xyz(1+y_2)}{z-(1-y)(1+y_2)}\, ,\quad
 y_b=\frac{z-(1-y)(1+y_2)}{z}\,,
\quad Q_b^2=Q^2z(1+y_2) \,.
\end{eqnarray}
More realistic (from the experimental point of view) is the calorimetric
event selection, when only the sum of the energies of the outgoing
electron and photon can be measured if the photon momentum lies inside
the small cone with opening angle $2\theta_0^{'}$ along the direction of
the final electron.  In this case we find
\begin{eqnarray}
\label{18}
\frac{z}{y}\frac{\dd^3\sigma^{\gamma\gamma}_{ii,cal}}{\dd x \, \dd y \, \dd z}
& = &
\frac{\alpha^2}{4\pi^2}P(z,L_0)\int\limits_{\Delta/Y}^{\infty}
\frac{\dd y_2}{(1+y_2)^3}\biggl[\frac{1+(1+y_2)^2}{y_2}(\widetilde L-1)+y_2
\biggr] \tilde \Sigma
\nonumber \\
& = &
\frac{\alpha^2}{4\pi^2}P(z,L_0)
\left[ (\widetilde L-1) \left(2 \ln \frac{Y}{\Delta} - \frac{3}{2} \right)
+ \frac{1}{2} \right] \tilde \Sigma \,.
\end{eqnarray}
In the last equation we used the relation
\begin{equation}
\Sigma_s=\frac{1}{(1+y_2)^2}\tilde \Sigma\,,
\end{equation}
which is valid for the calorimetric set-up.

Consider at last the semi-collinear region {\em iii)\/}. The relevant
contribution may be calculated using the quasireal electron
method~\cite{BFK}:
\begin{eqnarray}\label{19}
\frac{z}{y}\frac{\dd^3\sigma^{\gamma\gamma}_{iii}}{\dd x \, \dd y \, \dd z}
&=& \frac{\alpha}{2\pi}P(z,L_0)\frac{2\alpha}{\pi }
\int\frac{\dd^3 k_2}{\omega_2}\frac{\alpha^2(Q^2_{sc})}{Q^4_{sc}}I^{\gamma}\,,\\
\quad I^{\gamma}&=&B_{\rho\sigma}(zp_1,p_2,k_1)H^{\rho\sigma}/(8\pi)\,.
\nonumber
\end{eqnarray}
The quantity $B_{\rho\sigma}(zp_1,p_2,k_1)$ is obtained from equation
(\ref{8}), where it necessary to set $m = 0$. After some algebraic
transformations we obtain
\begin{eqnarray}
I^{\gamma} &=& \frac{1}{st}\Biggl[F_2(x_{sc},Q_{sc}^2)
\Biggl(\frac{M^2}{Q_{sc}^2}x_{sc}G
- V\Bigl[x_{sc}(z^2 + (1-y)^2) V + (1-y)(zQ^2 - s) -
\nonumber \\   \label{eq:21}
&&\phantom{\frac{1}{st}\Biggl[}- z(zQ^2 - t)\Bigr]\Biggr) - GF_1(x_{sc},Q_{sc}^2)\Biggr], \qquad
G = z^2Q^4 - 2st + Q_{sc}^4 \,,
\\  \nonumber
x_{sc} &=& \frac{Q_{sc}^2}{V(z+y-1)-2P\cdot k_2}\,, \quad
s = 2p_2\cdot k_2\,, \quad t = -2zp_1\cdot k_2\,, \quad
Q_{sc}^2 = zQ^2-s-t\,.
\end{eqnarray}
The angular integration in eq.~(\ref{19}) is to be performed over the
whole phase space, excepting the small cones along directions of motion
of the initial and scattered electrons that correspond to the kinematic
regions {\em i)} or {\em ii)}.  The result (for the details see
Appendix~B) has the form
\begin{eqnarray}\label{sig-iii}
\frac{z}{y}\frac{\dd^3\sigma^{\gamma\gamma}_{iii}}{\dd x \, \dd y \, \dd z} &=&
 \left(\frac{\alpha}{2\pi}\right)^2P(z,L_0)
 \Biggl[ \, \int\limits_{\Delta}^{x_2^t}
\frac{\dd x_2}{x_2}
\frac{z^2+(z-x_2)^2}{z(z-x_2)}\ln\frac{2(1-c)}{\theta_0^2}
\Sigma_t + \nonumber \\ 
&&
+\int\limits_{\Delta}^{x_2^s}\frac{\dd x_2}
{x_2}\frac{1+(1+y_2)^2}{1+y_2}\ln\frac{2(1-c)}{\theta_0^{'2}}
\Sigma_s + Z\Biggr] \,,\label{eq:22}\\[1em]
  \Sigma_t&=&\Sigma(x_t,y_t,Q_t^2)\,,
\end{eqnarray}
The logarithmic dependencies on the infrared regulator $\Delta$ and on
the angles $\theta_0$, $\theta_0'$ are fully contained in the first two
terms on the r.h.s., whereas the quantity $Z$ represents an integral
over the whole photon phase space of a well-behaved function, and it is
free from collinear and infrared singularities.  Its explicit expression
is given in Appendix~B.

The upper limits of the $x_2$-integration in (\ref{eq:22}) read
\begin{equation}\label{23}
x_2^t = z-\frac{Y(1+c)}{2} \, , \qquad x_2^s = \frac{2z-Y(1+c)}{1+c}\, ,
\end{equation}
and the arguments of $\Sigma_t$ are
\begin{equation} \label{24}
x_t = \frac{xy(z-x_2)}{z-x_2+y-1}\, , \quad y_t = \frac{z-x_2+y-1}{z-x_2}
, \quad Q_t = Q^2(z-x_2).
\end{equation}
An explicit expression for $x_m$, which is relevant for the calculation
of $Z$, is given in Appendix~B.

The formulae given above (see eqs.~(\ref{2}), (\ref{13}), (\ref{14}),
(\ref{16}) or (\ref{18}), and (\ref{eq:22}))
provide the complete answer for the leading and subleading contributions
up to the second order of perturbation theory.
The total sum of virtual, soft, and hard additional photons emission
corrections to the radiative DIS cross section does not depend on the
auxiliary parameter $\Delta=\delta\varepsilon/\varepsilon$, as it should
be.

\section{Results for different experimental situations}
\label{sez4}

The sum of the contributions of the leading and next-to-leading
corrections at order $\alpha^2$, which are given explicitly in
expressions (\ref{13}), (\ref{14}), (\ref{16}) or (\ref{18}), and
(\ref{eq:22}), may be written in the form
\begin{equation}\label{25}
\frac{z}{y} \frac{\dd^3\sigma}{\dd x \, \dd y \, \dd z}
=\left(\frac{\alpha}{2\pi}\right)^2 (\Sigma_i+\Sigma_f) \, .
\end{equation}
%
The first term $\Sigma_i$ is independent of the experimental selection
of the scattered electron and has the form
\begin{eqnarray}
\Sigma_i &=& \Biggl\{\frac{1}{2}L_0^2P^{(2)}_\Theta(z) +P(z,L_0)\Biggl[
\frac{1-16z-z^2}{2(1+z^2)}+\left(3-2\ln Y +\frac{4z}{1+z^2}\right)\ln z +
\nonumber \\ 
&&+ \ln^2 Y-2\Li(z)+2\Li\left(\frac{1+c}{2}\right)
-\frac{2(1+z)^2}{1+z^2}\ln (1-z)+
\frac{1-z^2}{2(1+z^2)}\ln^2 z \Biggr]\Biggr\} \tilde \Sigma +
\nonumber\\ \label{26}
&&+ P(z,L_0) \, \tilde \Sigma \, \ln\frac{2(1-c)}{\theta_0^2}
\Biggl[ \, \int\limits_{0}^{u_0}\frac{\dd u}{u}
(1+(1-u)^2)\Biggl(\frac{\Sigma_t}{(1-u)\tilde\Sigma}-1\Biggr) - \\ \nonumber
&&-
\int\limits_{u_0}^{1}\frac{\dd u}{u}(1+(1-u)^2)\Biggr]
+ P(z,L_0) \, Z, \qquad
u = \frac{x_2}{z}\ , \quad u_0 = \frac{x_2^t}{z} \,,
\end{eqnarray}
where $Z$ is given in Appendix~B and the remaining notations are as
above (see (\ref{14}), (\ref{eq:21}), and (\ref{23})).

The second term in (\ref{25}), denoted $\Sigma_f$, however, does
explicitly depend on the event selection.  It corresponds to the
emission of a hard photon by the scattered electron.  In the exclusive
set-up, when only the scattered bare electron is measured, while the
photon that is emitted close to the final electron's direction is
ignored, this contribution reads
\begin{eqnarray}
\Sigma_f = \Sigma_f^{excl} & = &
P(z,L_0)\int\limits_{0}^{x_2^s/Y}\dd y_2\Biggl[
\left(\frac{1+(1+y_2)^2}{y_2}(L_Q+\ln Y-1)+y_2\right)\frac{1}{1+y_2}\times
\nonumber \\ \label{27}
&&\times \Theta\left(y_2-\frac{\Delta}{Y}\right)
+ (L_Q+\ln Y-1)\delta(y_2)\left(2\ln\frac{\Delta}{Y}
+ \frac{3}{2}\right)\Biggr]\Sigma_s\,.
\end{eqnarray}
In this case the parameter $\theta'_0$, that separated the kinematic
regions {\em ii)} and {\em iii)}, only plays the role of an auxiliary
one; it has already cancelled in the above expression for the cross
section.

As we will see below, this situation is quite different for the
experimentally more realistic, calorimetric set-up, when the detector
cannot distinguish between events with a bare electron and events when
the electron is accompanied by a hard photon emitted within a small cone
with opening angle $2\theta_0^{'}$ around the direction of the scattered
electron.  For this case we obtain
\begin{eqnarray}
\Sigma_f=\Sigma_f^{cal} &=& P(z,L_0)\Biggl[\frac{1}{2}\tilde\Sigma
+ \ln\frac{2(1-c)}{\theta_0^{'2}}\int\limits_{0}^{\infty}\frac{\dd y_2}{y_2}
\frac{1+(1+y_2)^2}{1+y_2}\times \nonumber \\
&&\times \biggl(\Sigma_s\Theta(y_2^{\mathrm{max}}-y_2)
- \frac{\tilde\Sigma}{(1+y_2)^2}\biggr)\Biggr]\,.
\end{eqnarray}
For the calorimetric event selection the parameter $\theta_0^{'}$ is a
physical one and the final result therefore does depend on it.  However,
the mass singularity that is connected with the emission of the photon
off the scattered electron is cancelled in accordance with the
Kinoshita-Lee-Nauenberg theorem~\cite{KLN}.

Note that the case of a coarse detector for the scattered electron,
i.e., $\theta_0^{'} \sim {\mathcal O}(1)$, agrees at the level of leading
logarithms with the result of paper \cite{AAKM:ll}, that was obtained in
the approximation of absence of emission along the scattered electron.
Our result disagrees with the result of Bardin et al.~\cite{bar} on the
radiative corrections, as they neglected the interference of the
emission of two photons; see \cite{AAKM:ll} for a detailed discussion.

Finally we will give some numerical values obtained for the radiative
corrections at leading and next-to-leading order for the experimentally
relevant case of the calorimetric event selection with a realistic
resolution.  As input we used
\begin{equation}
 E_e = 27.5 \mbox{ GeV}\,, \quad
 E_p = 820 \mbox{ GeV}\,, \quad
 \theta_0 = 0.5 \mbox{ mrad}\,,
\end{equation}
whereas for the resolution of the detector we assumed $\theta_0' = 50
\mbox{ mrad}$.  As structure function we chose the ALLM97
parameterization \cite{ALLM97} with $R=0$.  No cuts were applied to the
photon phase space.

\begin{figure}
\begin{center}
\begin{picture}(100,100)
\put(0,0){
\includegraphics[height=100mm]{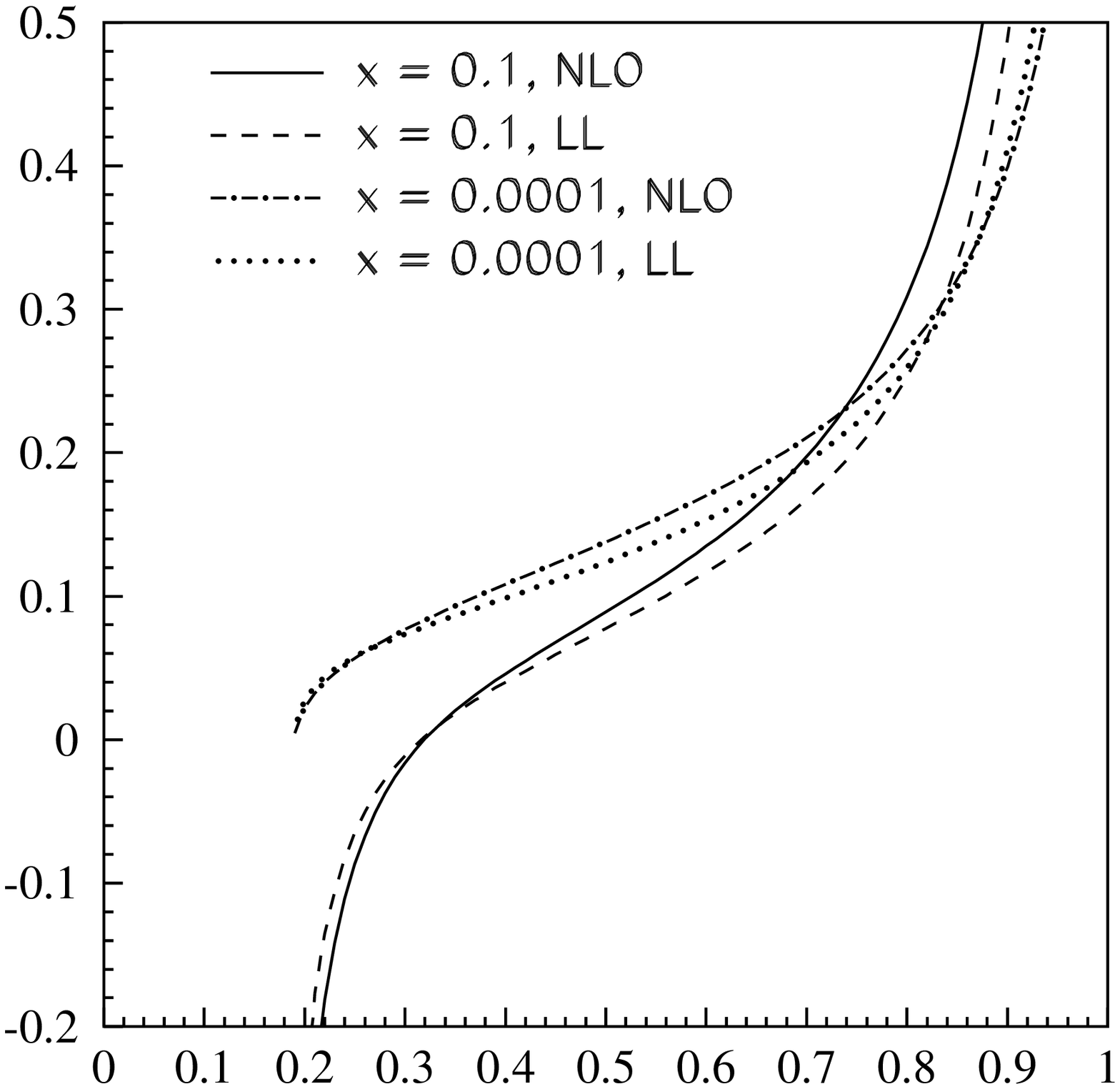}
}
\put(86,-1){$y$}
\put(-5,83){$\delta_\matrm{RC}$}
\end{picture}
\end{center}
\caption{Radiative corrections $\delta_\matrm{RC}$
(\protect\ref{eq:delta}) at leading and next-to-leading order for
standard Bjorken variables for $x=0.1$ and $x=10^{-4}$ and a tagged
photon energy of 5~GeV.}
\label{fig1}
\end{figure}

Figure~\ref{fig1} compares the radiative correction
\begin{equation} \label{eq:delta}
        \delta_\mathrm{RC} =
        \frac{\dd^3\sigma}{\dd^3\sigma_\mathrm{Born}} - 1
\end{equation}
at leading and next-to-leading order for the measurement in terms of the
standard Bjorken variables $x$ and $y$ for $x=0.1$ and $x=10^{-4}$ for a
tagged energy of $E_\mathrm{PD} =  5$~GeV.  The typical size of the
next-to-leading order contributions with respect to the
leading-logarithmic result \cite{AAKM:ll} amount to up to the order of
5\%, depending on the kinematic region.  The apparent cutoff in the
curves at low $y$ appears at
\begin{equation}
        y_\mathrm{Bj,min} = \frac{1-z}{1-xz}
\end{equation}
due to the condition that $\hat{x} \leq 1$, see eq.~(\ref{4}).

\begin{figure}
\begin{center}
\begin{picture}(100,100)
\put(0,0){
\includegraphics[height=100mm]{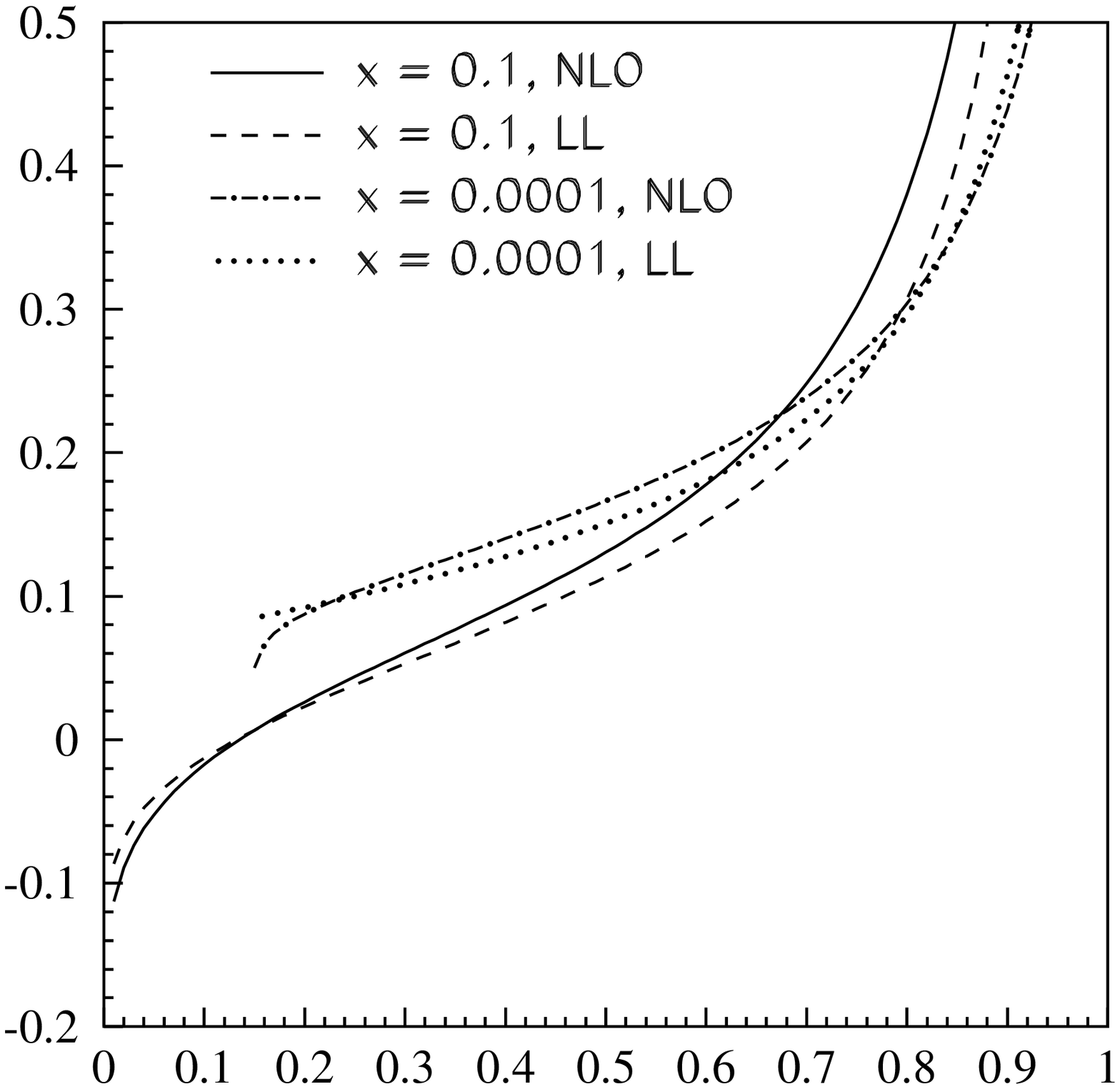}
}
\put(86,-1){$\hat{y}$}
\put(-5,83){$\delta_\matrm{RC}$}
\end{picture}
\end{center}
\caption{Radiative corrections $\delta_\matrm{RC}$
(\protect\ref{eq:delta}) at leading and next-to-leading order for
shifted Bjorken variables for $\hat{x}=0.1$ and $\hat{x}=10^{-4}$ and a
tagged photon energy of 5~GeV.}
\label{fig2}
\end{figure}

For the case of the shifted Bjorken variables (\ref{eq:kin-vars}) we
present the corresponding result in fig.~\ref{fig2}.  Here no lower cut in
$\hat{y}$ appears, like in the case of standard Bjorken variables.
However, the curves for small $\hat{x}=10^{-4}$ are not continued below
the value of $\hat y$ where the cones with opening angles $2\theta_0$
(PD) and $2\theta_0'$ (calorimeter resolution) would start to overlap,
as $\hat y \to 0$ corresponds to forward scattering.  The difference of
the radiative corrections between leading and next-to-leading order in
this case also amounts to the order of 5\%.

\begin{figure}
\begin{center}
\begin{picture}(100,100)
\put(0,0){
\includegraphics[height=100mm]{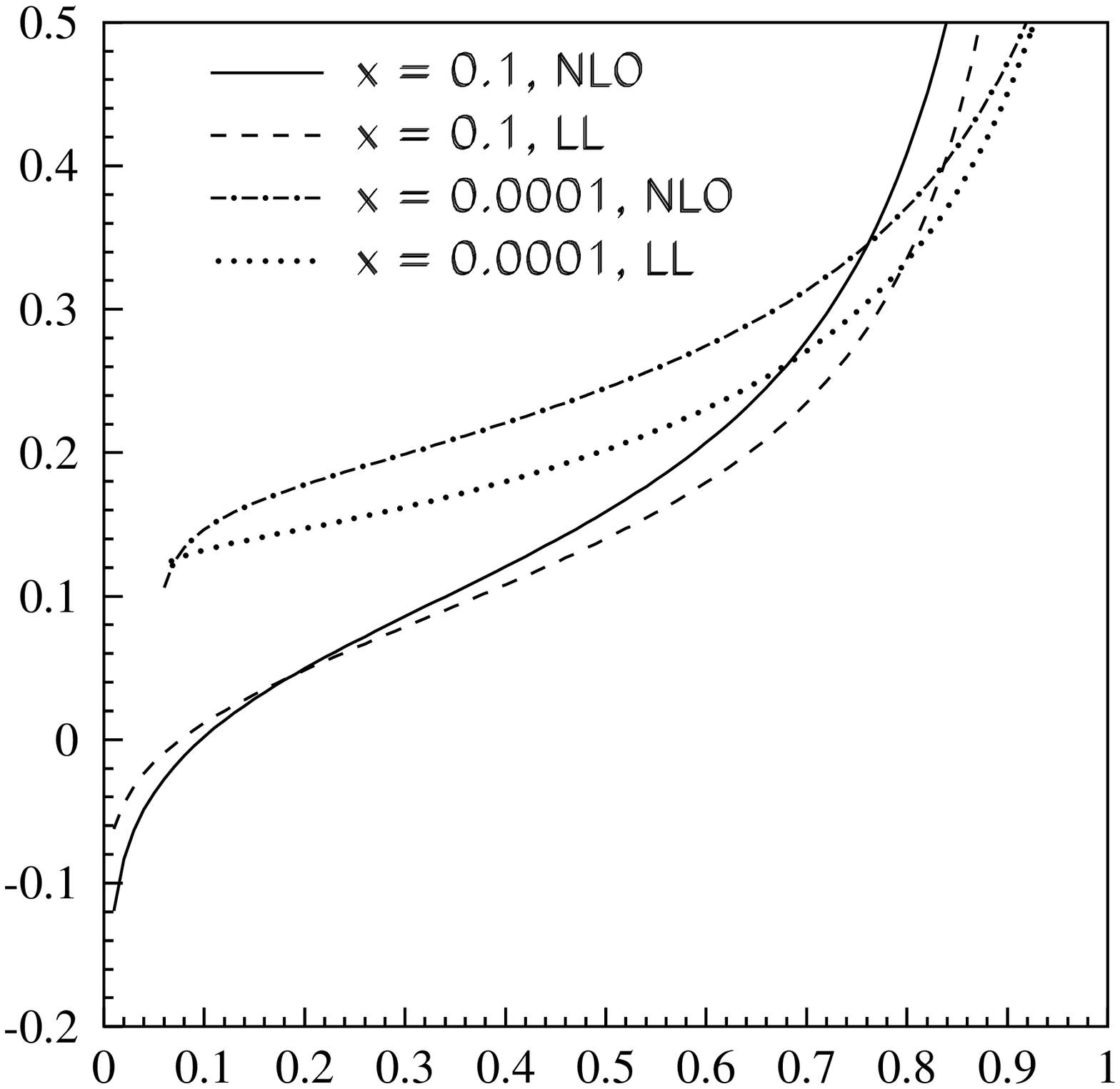}
}
\put(86,-1){$\hat{y}$}
\put(-5,83){$\delta_\matrm{RC}$}
\end{picture}
\end{center}
\caption{Radiative corrections $\delta_\matrm{RC}$
(\protect\ref{eq:delta}) at leading and next-to-leading order for
shifted Bjorken variables for $\hat{x}=0.1$ and $\hat{x}=10^{-4}$ and a
tagged photon energy of 20~GeV.}
\label{fig3}
\end{figure}

Lastly, in order to exhibit the $z$-dependence of the next-to-leading
order contributions, i.e., the dependence on the energy in the photon
tagger, we plot the corrections for $E_\mathrm{PD} = 20$~GeV in fig.~\ref{fig3}
for $\hat{x} = 0.1$ and $\hat{x}=10^{-4}$.  One sees in particular the
increasing relevance of the next-to-leading order terms for large
$E_\mathrm{PD}$ and in the experimentally interesting range of small
$\hat{x}$.

We note in conclusion that the set of Feynman diagrams considered here
is gauge invariant and model independent but not complete.  We have neglected
the contributions with two virtual photons exchanged between electron and
the target that appear at the same order of perturbation theory, as well
as the interference with the contributions when the second photon is
emitted by the hadronic side.  (These contributions have not yet been
considered in any literature about DIS known to the authors).  However,
the description of this part is definitely model dependent and will be
discussed elsewhere.

\subsection*{Acknowledgments}

The authors are indebted to D.~Bardin, E.~Boos, L.~Kalinovskaya, T.~Ohl,
T.~Riemann, and L.~Trentadue for useful discussions and interest in the
problem considered.  H.A. would like to thank L. Favart, M. Fleischer,
and S. Schleif for continued interest.  This work was supported in part
by INTAS grant 93--1867 ext., and the Heisenberg-Landau program.



\section*{Appendix A \\ List of angular integrals for virtual corrections}

In this section we collect the results of the angular integration of the
definite structures of the Compton tensor in \cite{compt}.  Using
integrals similarly to (\ref{9}) and retaining only terms that contain
at least one large logarithm $L_0$ or $L_Q$, we obtain
\begin{eqnarray}
\frac{2\varepsilon^2}{Q_l^2}\int\frac{\dd \Omega_k}{2\pi} T_g
&=& - \rho \biggl[\frac{1+z^2}{(1-z)^2}(L_0-1) + 1 \biggr]
+ \frac{1+z^2}{(1-z)^2}[ A\ln z + B ]
- \frac{4z}{(1-z)^2}L_Q\ln z - \nonumber \\ \nonumber
&&- \frac{2-(1-z)^2}{2(1-z)^2}L_0\,, \\ \nonumber
2\varepsilon^2 \int\frac{\dd \Omega_k}{2\pi} T_{11}
&=& \frac{4z}{(1-z)^2} \rho L_0 - \frac{2z(1+(1-z)^2)}{(1-z)^4}
( A\ln z + B ) - A \frac{z(3-z)}{(1-z)^3} + \\ \nonumber
&&+ \frac{2L_0}{(1-z)^3} \biggl(
\frac{z(8z-3)}{1-z}\ln z + 2z + z^2 \biggr)\,, \\
2\varepsilon^2 \int\frac{\dd \Omega_k}{2\pi} T_{22}
&=&  \rho\biggl(\frac{4z}{(1-z)^2} L_0 - \frac{8}{(1-z)^2} \biggr)
+ \frac{16}{(1-z)^2} \ln z L_Q - \\ \nonumber
&&- \frac{2z(1+2(1-z)^2)}{(1-z)^4}( A\ln z + B )
- A \frac{3z-1}{z(1-z)^3} +\\ \nonumber
&&+ \frac{2L_0}{z(1-z)^3}\biggl(
\frac{1+4z(z^2+z-1)}{1-z}\ln z + z^3 - z^2 + 4z - 1 \biggr)\,, \\ \nonumber
2\varepsilon^2 \int\frac{\dd \Omega_k}{2\pi} T_{21}
&=&  \frac{2z^2}{(1-z)^4} ( A\ln z + B )
+ \frac{3z-1}{(1-z)^3}A +\\ \nonumber
&&+ \frac{2L_0}{(1-z)^3}\biggl(
\frac{-1+4z-4z^2-4z^3}{1-z}\ln z
- 2z^2 - 2z + 1 \biggr)\,, \\ \nonumber
2\varepsilon^2 \int\frac{\dd \Omega_k}{2\pi} T_{12}
&=& \frac{2z(2-z)}{(1-z)^4}( A\ln z + B )
+ \frac{3-z}{(1-z)^3}A + \frac{2L_0}{(1-z)^3}\biggl(
\frac{3-8z}{1-z}\ln z - 1 - 2z \biggl)\,,
\end{eqnarray}
where $\rho$, $A$ and $B$ are given by eq.~(\ref{10}).
It is remarkable to see that the relation
\begin{equation}
\int\limits \frac{\dd \Omega_k}{2\pi}
\left[ 4zT_g + Q_l^2(T_{11}+z^2T_{22}+zT_{12}+zT_{21}) \right] = 0
\end{equation}
is fulfilled, leading to the factorization of the virtual corrections in
eq.~(\ref{10}).

\section*{Appendix B \\
Angular integration for bremsstrahlung corrections}

To perform the angular integration in (\ref{19}) we first represent
the integrand in the form
\begin{equation}
\label{eq:F}
\frac{\eps^2 \alpha^2(Q_{sc}^2)I^{\gamma}}{Q_{sc}^4}
= \frac{F(t_1,t_2)}{t_1t_2}\, ,
\end{equation}
where $t_{1,2}=(1-c_{1,2})/2$, $c_{1,2}=\cos\theta_{1,2}$, and
$\theta_{1,2}$ are the angles between non-tagged photon momentum $k_2$
and the momenta of the initial and the scattered electrons.
Note that $F(t_1,t_2)$ behaves regularly for $t_1 \to 0$ or $t_2 \to 0$.
This can be easily seen by considering the limiting cases for the
quantity $I^{\gamma}$.

For the case $t\to 0$, which corresponds to the the second photon being
emitted close to the direction of the incoming electron, one obtains
from eq.~(\ref{eq:21})
\begin{eqnarray}
I^{\gamma}|_{t \to 0}&=&\frac{Q^2}{x_2t}(z^2+(z-x_2)^2)\times\nonumber\\
&&\times\left[x_tF_2(x_t,Q_t^2)
\biggl(\frac{M^2}{Q_t^2}-\frac{1-y}{x^2y^2(z-x_2)}\biggr)-F_1(x_t,Q_t^2)
\right],
\end{eqnarray}
while for the case $s \to 0$, corresponding to the second photon being
almost collinear to the final electron,
\begin{eqnarray}
I^{\gamma}|_{s\to 0}&=&-\frac{Q^2z}{y_2s}(1+(1+y_2)^2)\times\nonumber\\
&&\times\left[x_bF_2(x_b,Q_b^2)
\biggl(\frac{M^2}{Q_b^2}-\frac{1-y}{x^2y^2z(1+y_2)}\biggr)-F_1(x_b,Q_b^2)
\right],
\end{eqnarray}
see eqs.(\ref{eq:17}) and (\ref{24}) for the notation.  The r.h.s.\ of
eq.~(\ref{eq:F}) is easily seen to be
\begin{eqnarray}
\left. \frac{\eps^2 \alpha^2(Q^2_{sc}) I^{\gamma}}{Q^4_{sc}}
\right|_{t \to 0} & = &
\frac{1}{t_1t_2} \frac{a}{16\pi x_2^2} \frac{z^2+(z-x_2)^2}{z(z-x_2)}
\, \Sigma(x_t,y_t,Q^2_t)\, , \\
\left. \frac{\eps^2 \alpha^2(Q^2_{sc}) I^{\gamma}}{Q^4_{sc}}
\right|_{s \to 0} & = &
\frac{1}{t_1t_2} \frac{a}{16\pi x_2^2} \frac{1+(1+y_2)^2}{1+y_2}
\, \Sigma(x_b,y_b,Q^2_b)\,, 
\end{eqnarray}
where $a = (1 - \cos\theta)/2$.

For the phase space of the photon we use the following representation:
\begin{eqnarray}
&& \int\limits_{}^{}\frac{\dd^3 k_2}{\omega_2} =
\varepsilon^2\int\limits x_2\,\dd x_2 \,\dd \Omega_2 =
 4\varepsilon^2 \int\limits_{}^{}x_2\,\dd x_2
\int \frac{\dd t_1\,\dd t_2}{\sqrt{D}}\Theta(D)\,, \\ \nonumber
&& D=(t_2-y_-)(y_+-t_2),\quad
y_{\pm}=t_1(1-2a)+a\pm2\sqrt{a(1-a)t_1(1-t_1)} \,.
\end{eqnarray}
The region of integration is determined by the conditions
\begin{equation}
\sigma_1<t_1<1, \quad \sigma_2<t_2<1, \quad D>0, \quad \sigma_1=\frac{\theta_0^2}{4},
\quad \sigma_2=\frac{\theta_0^{'2}}{4}.
\end{equation}
Using the substitution
\begin{equation}
t_2\to t_2(t_1,u)=\frac{(a-t_1)^2(1+u^2)}{y_++u^2y_-}\, ,
\end{equation}
and the identity
\begin{eqnarray}
\int\limits_{\sigma_1}^{1}\dd t_1\int\limits_{\sigma_2}^{1}\dd t_2
\frac{F(t_1,t_2)}{t_1t_2\sqrt{D}}\Theta(D)
&=& \frac{\pi}{a}\left[ F(a,0)\ln\frac{a}{\sigma_2}
+ F(0,a)\ln\frac{a}{\sigma_1} \right] + \nonumber \\
&& + 2 \int\limits_{0}^{\infty}\frac{\dd u}{1+u^2}\lim_{\eta\to 0}
\Biggl[\int\limits_{\eta}^{1}\frac{\dd t_1}{t_1|t_1-a|}
\left(F(t_1,t_2)-F(a,0)\right) + \nonumber\\
&&\phantom{+ 2 \int\limits_{0}^{\infty}}
+\int\limits_{\eta}^{a}
\frac{\dd t_1}{t_1a}\left(F(a,0)-F(0,a)\right)\Biggr]\,,
\end{eqnarray}
which is valid for $\sigma_1, \sigma_2 \ll a$, we obtain for $Z$
from eq.~(\ref{eq:22}) the following expression:
\begin{eqnarray}
Z&=&-\frac{4(1-c)}{zQ^2}\int\limits_{0}^{\infty}\frac{\dd u}{1+u^2}
\Biggl[\int\limits_{\eta}^{1}
\frac{\dd t_1}{t_1|t_1-a|}\int\limits_{0}^{x_m}\frac{\dd x_2}{x_2}
(\Phi(t_1,t_2(t_1,u))-\Phi(a,0)) + \nonumber\\ 
&&\phantom{-\frac{4(1-c)}{zQ^2}\int\limits_{0}^{\infty}\frac{\dd u}{1+u^2}\Biggl[}
+ \int\limits_{\eta}^{a}\frac{\dd t_1}{t_1a}\int\limits_{0}^{x_m}
\frac{\dd x_2}{x_2}(\Phi(a,0)-\Phi(0,a))\Biggr]_{\eta\to 0}\,,
\end{eqnarray}
where
\begin{eqnarray}
\Phi(t_1,t_2) &=& \frac{\alpha^2(Q^2_{sc})stI^{\gamma}}{Q^4_{sc}}
\Bigg|_{c_1=1-2t_1,\ c_2 = 1-2t_2(t_1,u),\ c = 1 - 2a}.
\end{eqnarray}
The upper limit of the $x_2$-integration, $x_m$, may be deduced from
\cite{bar}.  It has the form
\begin{eqnarray}
x_m &=& \frac{z(e+p)-\Delta_m-Y(e+z)-(p-z)Yc}{z+e-Y+(p-z)c_1+Yc_2},
\qquad e=\frac{E_p}{\varepsilon}\,, \nonumber \\
p &=& \frac{P_p}{\varepsilon}\,, \qquad
\Delta_m=\frac{(M+m_{\pi})^2-M^2}{2\varepsilon^2}\,.
\end{eqnarray}
This finally leads to eq.~(\ref{sig-iii}).

It is important to note that when calculating $Z$ one encounters neither
collinear nor infrared singularities.


\end{document}